\documentclass[useAMS,usenatbib,longnamesfirst,usedcolumn]{mn2e} 
\usepackage{epsfig}	
\usepackage{times}	
\title[Cluster gas properties and cool-core/non-cool core bimodality]{A statistically-selected Chandra 
 sample of 20 galaxy clusters -- II. Gas properties and cool-core/non-cool core bimodality}
\author[A. J. R. Sanderson, E. O'Sullivan and T. J. Ponman]
       {Alastair J. R. Sanderson$^{1}$\thanks{E-mail: ajrs@star.sr.bham.ac.uk},
        Ewan O'Sullivan$^{2}$ and Trevor J. Ponman$^{1}$\\
 $^{1}$School of Physics and Astronomy, University of
        Birmingham, Edgbaston, Birmingham B15 2TT, UK \\
 $^{2}$Harvard-Smithsonian Center for Astrophysics, 60 Garden Street,
        Cambridge, MA 02138\\
       \\}
 \date{Accepted 2009 February 09.
      Received 2009 February 09;
      in original form 2008 November 25 ($svn$ $Revision: 136 $)}

\pagerange{\pageref{firstpage}--\pageref{lastpage}}
\pubyear{2008}

\shortcites{birzan04,baldi07,balogh01,bauer05,blanton04,croston08,
degrandi04,donahue06,donahue05,finoguenov01,ikebe02,kay07,mccarthy08,
mukherjee98,peterson01,piffaretti05,poole08,poole06,pratt07,rasera08,
rebusco05,sun07,sun09,vikhlinin06,vikhlinin05,voigt02,voit02,voit08,
zhang06}

\defcitealias{san06}{Paper~I}

\newcommand{\rmsub}[2]{\ensuremath{#1_{\mathrm{#2}}}} 
\newcommand{\srel}[2]{\mbox{\ensuremath{#1 - #2}}} 
\newcommand{\BeppoSAX}{\textit{BeppoSAX}}
\newcommand{\Chandra}{\textit{Chandra}}
\newcommand{\chisq}{\ensuremath{\chi^2}}
\newcommand{\CIAO}{\textsc{ciao}}
\newcommand{\cm}{\ensuremath{\mbox{~cm}}}
\newcommand{\cmsq}{\ensuremath{\cm^2}}
\newcommand{\keV}{\ensuremath{\mbox{~keV}}}
\newcommand{\km}{\ensuremath{\mbox{~km}}}
\newcommand{\kmpspMpc}{\ensuremath{\km \ps \pMpc\,}}
\newcommand{\kpc}{\ensuremath{\mbox{~kpc}}}
\newcommand{\Mpc}{\ensuremath{\mbox{~Mpc}}}
\newcommand{\MT}{\srel{M}{\TX}}
\newcommand{\omegal}{\rmsub{\Omega}{\Lambda}}
\newcommand{\omegam}{\rmsub{\Omega}{m}}
\newcommand{\pMpc}{\ensuremath{\Mpc^{-1}}}
\newcommand{\ps}{\ensuremath{\s^{-1}}}
\newcommand{\rtwoh}{\rmsub{r}{200}}
\newcommand{\rfiveh}{\rmsub{r}{500}}
\newcommand{\rhogas}{\rmsub{\rho}{gas}}
\newcommand{\Rproject}{\textsc{r}}
\newcommand{\s}{\ensuremath{\mbox{~s}}}
\newcommand{\Tbar}{\ensuremath{\overline{T}}}
\newcommand{\TX}{\rmsub{T}{X}}
\newcommand{\XMM}{\emph{XMM-Newton}}
\newcommand{\XSPEC}{\textsc{xspec}}
\begin{document}

\maketitle

\label{firstpage}

\begin{abstract} \noindent We investigate the thermodynamic and 
chemical structure of the intracluster medium (ICM) across a 
statistical sample of 20 galaxy clusters analysed with the \Chandra\ 
X-ray satellite. In particular, we focus on the scaling properties of 
the gas density, metallicity and entropy and the comparison between 
clusters with and without cool cores (CCs). We find marked differences 
between the two categories except for the gas metallicity, which 
declines strongly with radius for all clusters ($Z\propto r^{-0.31}$), 
outside $\sim$0.02\rfiveh. The scaling of gas entropy is 
non-self-similar and we find clear evidence of bimodality in the 
distribution of logarithmic slopes of the entropy profiles. With only 
one exception, the steeper sloped entropy profiles are found in CC 
clusters whereas the flatter slope population are all non-CC clusters. 
We explore the role of thermal conduction in stabilizing the ICM and 
conclude that this mechanism alone is sufficient to balance cooling in 
non-CC clusters. However, CC clusters appear to form a distinct 
population in which heating from feedback is required in addition to 
conduction. Under the assumption that non-CC clusters are thermally 
stabilized by conduction alone, we find the distribution of Spitzer 
conduction suppression factors, $f_c$, to be log-normal, with a log 
(base 10) mean of $-1.50\pm0.03$ (i.e. $f_c=0.032$) and log standard 
deviation $0.39\pm0.02$.

\end{abstract}



\section{Introduction}
\label{sec:intro}
The majority of baryons in collapsed massive haloes reside in a hot phase,
in the form of a gaseous intracluster medium (ICM), with the remainder
predominantly locked up in stars \citep{fukugita98,gonzalez07}. This hot
gas serves as a reservoir of material to fuel not only star formation, but
also black hole growth, as the ultimate endpoints of radiative
cooling. Both these processes in turn give rise to feedback from subsequent
supernova winds \citep{katz92,strickland00} and outbursts from active
galactic nuclei \citep[AGN; see][for a recent review]{mcnamara07},
respectively.

Given the considerable effectiveness of radiative cooling in depleting the
hot gas, its dominance of the baryon budget in cluster haloes emphasizes
the importance of feedback in order to restrict the excessive growth of
galaxies \citep[e.g.][]{cole91} and avoid a `cooling crisis'
\citep{balogh01}--- which has long plagued cosmological simulations in
which the effects of non-gravitational heating are neglected
\citep[e.g.][]{katz93,suginohara98}. The same feedback mechanism(s) may also
be responsible for arresting gas cooling in dense cluster cores
\citep[e.g.][]{peterson01}, where the development of a classical `cooling
flow' \citep{fabian94b} appears to be truncated \citep{peterson06}.

A further indication of the importance of cooling is the discovery of short
gas cooling times in the inner regions of even non-cool core clusters
\citep{san06}, as well as the quasi-universality of cooling time profiles
across a wide range of cluster masses \citep{voigt04,bauer05,san06},
despite the clear differences between the temperature profiles of cool-core
and non-cool core clusters \citep[e.g.][]{san06,zhang06,pratt07}. The
explanation for this dichotomy in the cluster population is the subject of
current debate \citep[e.g.][]{mccarthy08,guo08}, but is likely to involve
galaxy feedback, given that cosmological simulations appear to over-predict
the abundance of cool-cores in the cluster population \citep[e.g.][]{kay07}
and that there is some question over the effectiveness of merging in
permanently erasing cool cores \citep{poole06}. However, the lack of
evidence for \emph{strong} shock heating from most AGN outbursts in cluster
cores \citep{mcnamara07} presents a challenge to understanding how feedback
alone is sufficient to offset cooling losses.

It is clear that a more complete picture of the thermodynamic state of hot
gas in clusters is needed, in order both to solve the cooling flow problem
and tackle the broader issue of feedback between galaxies and the
intracluster medium. In pursuing these goals it is necessary to map the
thermodynamic state of the ICM across a wide mass range, including both
cool core and non-cool core clusters, the latter of which are known to be
under-represented in such detailed studies. This enables the gas entropy to
be determined, which is a very sensitive probe of non-gravitational
processes, as well the metallicity, which acts as a tracer of supernova
enrichment and gas mixing. We aim to do this using \Chandra\ observations
of nearby clusters, which is the only instrument able to reliably probe
core gas properties on kpc scales, where the effects of baryon physics are
greatest. The basis for this investigation is the statistical sample of
\citet[hereafter \citetalias{san06}]{san06} comprising 20 galaxy clusters, 
in order to provide a representative survey of detailed cluster properties
in the local universe.

Throughout this paper we adopt the following cosmological parameters:
$H_{0}=70$\kmpspMpc, $\rmsub{\Omega}{m}=0.3$ and
$\Omega_{\Lambda}=0.7$. Throughout our spectral analysis we have used
\XSPEC\ 11.3.2t, incorporating the solar abundance table of
\citet{grevesse98}, which is different from the default abundance
table. Typically this results in larger Fe abundances, by a factor of
$\sim$1.4. All errors are 1$\sigma$, unless otherwise stated.

\section{Sample selection}
The objects studied in this paper comprise the statistical sample of 20
galaxy clusters observed with \Chandra\ presented in \citetalias{san06}.
The sample contains the 20 highest flux clusters drawn from the 63 cluster,
flux-limited sample of \citet{ikebe02}, excluding those objects with
extremely large angular sizes (the Coma, Fornax and Centaurus clusters),
which are difficult to observe with \Chandra\ owing to its limited field of
view. The \citeauthor{ikebe02} flux-limited sample was itself constructed
from the HIFLUGS sample of \citet{reiprich02}, additionally selecting only
those clusters lying above an absolute galactic latitude of 20 degrees and
located outside of the Magellanic Clouds and the Virgo Cluster regions.

\subsection{Re-analysis of Chandra data}
Since the original analysis of the statistical sample data in 
\citetalias{san06}, all but one of the \Chandra\ observations have been 
reprocessed with uniform calibration by the \Chandra\ X-ray Center (CXC)
and we have reanalysed all the data accordingly. The only exception is
Abell~2256 (ObsID 1386), which was observed at a frontend temperature of
-110\,C: thus far, only datasets observed at -120\,C have been reprocessed, in
order to provide a uniform calibration of most of the data in the \Chandra\
archive\footnote{for details, see
http://cxc.harvard.edu/cda/repro3.html}. Despite this, there is no
indication that our results for Abell~2256 are anomalous in any way.

For the clusters Abell~401, Abell~496, Abell~1795, Abell~2142 and
Abell~4038, longer observations are now available and, in the case of
Abell~1795 \& Abell~2142 these are with ACIS-I, giving the advantage of a
wider field-of-view over the datasets analysed in \citetalias{san06}. For
Abell~478, however, we retain our original ACIS-S analysis (ObsID 1689) in
favour of a newer ACIS-I observation, as the latter exposure is much
shallower. Details of the new datasets analysed are given in
Table~\ref{tab:bcc_new} and key properties for the full sample are listed
in Table~\ref{tab:main}. The data analysis and reduction were performed as
detailed in \citetalias{san06}, using version 3.4 of the standard software
--- \Chandra\ Interactive Analysis of Observations
(\CIAO\footnote{http://cxc.harvard.edu/ciao/}), incorporating
\textsc{caldb} version 3.4.2.

%
\begin{table}
\centering
\begin{tabular}{l*{3}{c}}
\hline \\[-2ex]
Name & Obsid$^{a}$ & Detector$^{b}$ & Datamode$^{c}$ \\
\hline\\[-2ex] 
 Abell 401  & 2309 & I & F \\
 Abell 496  & 4976 & S & VF \\
 Abell 1795 & 5289 & I & VF \\
 Abell 2142 & 5005 & I & VF \\
 Abell 4038 & 4992 & I & VF \\ \\[-2ex]
\hline
\end{tabular}
\caption{
 Clusters from the statistical sample for which new observations 
 have been analysed. $^{a}$Chandra observation identifier. $^{b}$Denotes either
 ACIS-I or ACIS-S. $^{c}$Telemetry data mode (either Faint or Very faint).}
\label{tab:bcc_new}
\end{table}

\section{Data analysis}
Spectral fitting was performed as described in \citet{san06}, using
weighted response matrix files (RMFs) generated with the \CIAO\ task
'\textsc{mkacisrmf}' for all cases, except Abell~2256 -- the only
non-reprocessed dataset -- where the older task `\textsc{mkrmf}' was used
instead. Spectra were fitted using an absorbed (\textsc{wabs}
\textsc{xspec} component) \textsc{apec} model over the energy range 
0.5--7.0 keV for observations made with the ACIS-S detector, and 0.7--7.0
keV for those made with ACIS-I (as indicated in column~3 of
Table~\ref{tab:main}). Spectra were grouped to a minimum of 20 counts per
bin and fitted using the \chisq\ statistic.

\subsection{Cluster mean temperature and fiducial radius}
\label{sec:mean_kT}
Measurements of mean temperature, \Tbar, and fiducial scaling radii are
very important in scaling studies, in order to permit fair a comparison of
properties across a wide range of cluster mass. In \citetalias{san06} we
used an iterative scheme to determine both a core-excluded mean temperature
and \rfiveh\ (the radius enclosing a mean overdensity of 500 with respect
to the critical density of the Universe), via the \MT\ relation of
\citet{finoguenov01}. However, we have subsequently refined this method in 
the following two ways, to improve the reliability of our results.

Firstly, we have adopted the newer, \Chandra-derived \MT\ relation of
\citet{vikhlinin06} which is based on clusters with high quality 
observations, allowing direct measurements of gas properties at \rfiveh.
Secondly, we have now excluded a larger central region of the data
(0.15\rfiveh), to remove more completely any contaminating emission from
strong central cooling that may occur in the core. Our chosen annular
extraction region therefore spans the range 0.15--0.2\rfiveh; we use
0.2\rfiveh\ to exclude outer regions where data incompleteness begins to
affect our sample, due to the restricted \Chandra\
field-of-view. None the less, one advantage of using an outer annulus of
0.1--0.2\rfiveh\ is that it provides a better measure of any central
temperature drop (within 0.1\rfiveh), as this range typically brackets the
peak around the cool core radius. Therefore we retain the objective
classifications of cool-core status determined in
\citetalias{san06} (and listed in Table~\ref{tab:main}), which were based
on the significance of the temperature difference between the spectrum
extracted in the range 0.1--0.2\rfiveh\ compared to that measured inside
0.1\rfiveh.

The temperatures in the \MT\ relation of \citet{vikhlinin06} were based on
an aperture extending out to \rfiveh, excluding the central 70\,kpc. Since
cluster temperature profiles generally decline monotonically with radius
outside of any cool core \citep[e.g.][]{vikhlinin05,pratt07}, it is
therefore necessary to allow for a bias factor that would shift temperature
measurements within 0.2\rfiveh\ higher compared to those made within
\rfiveh. This was done by comparing the temperatures obtained using the
iterative method of \citetalias{san06} in the range 0.15--0.2\rfiveh\ for 7
clusters from the sample of \citet{vikhlinin06} which were analysed in this
work and in a companion analysis of galaxy groups (O'Sullivan et al. in
prep.), namely Abell 262, 478, 1795, 2029 and MKW~4, MS1157 and
USGCS152. The aim of this process was to determine a multiplicative factor
to apply uniformly to the temperatures measured in the range
0.15--0.2\rfiveh\ to bring them into line with those measured in the
\citeauthor{vikhlinin06} aperture. The determination of this rescale factor
was also iterative, and the procedure was as follows.

Firstly, a mean temperature (\Tbar) was measured iteratively in the range
0.15--0.2\rfiveh\ using the \citeauthor{vikhlinin06} \MT\ relation to
achieve convergence for each of the 6 clusters common to both samples. Then
a rescale factor was calculated as the mean ratio of these temperatures to
the corresponding ones quoted in \citet{vikhlinin06}. Secondly, the
determination of \Tbar\ in the range 0.15--0.2\rfiveh\ was repeated, using
this rescale factor to adjust the measured temperature before using
Equation~\ref{eqn:r500} to calculate \rfiveh. The process was repeated
until convergence of the rescale factor, $f$, whose final value was
determined to be 0.96, i.e. temperatures measured in our aperture were 4
per cent hotter than those measured by \citeauthor{vikhlinin06}. For a
cluster of redshift, $z$, the radius is given by
\begin{equation}
\label{eqn:r500}
\rfiveh = \frac{f\times 484.7\times \Tbar^{0.527}}{E(z)}\,\, \mathrm{kpc}
\end{equation} 
where $f=0.96$ is a rescale factor to convert temperatures measured in the
range 0.15--0.2\rfiveh\ with those in the range 70\,kpc--\rfiveh, and
\begin{equation}
 E(z) = (1+z) \sqrt{1+(z\,\,\omegam)+\frac{\omegal}{(1+z)^2}-\omegal}\,\,.
\end{equation}

\begin{figure}
\includegraphics[width=8.3cm]{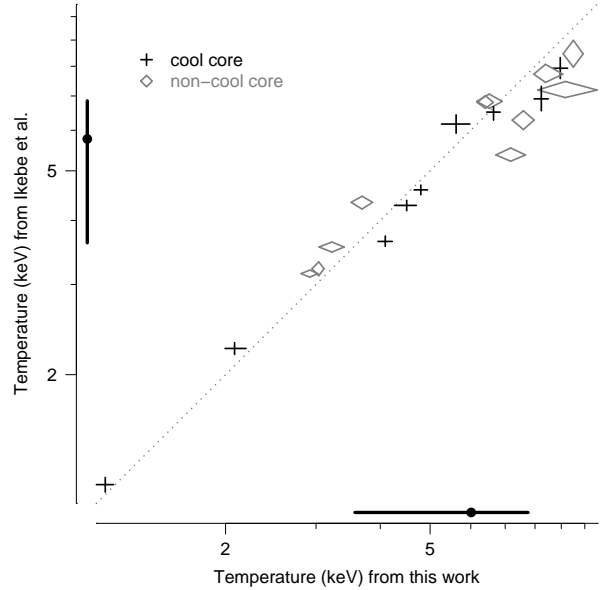}
\caption{ \label{fig:compare_kT}
 A comparison of the new mean temperatures from this work
 (0.15--0.2\rfiveh) and those of \citet{ikebe02}, showing the line of
 equality. The dots and solid lines indicate the marginal medians and
 interquartile ranges.}
\end{figure}

A comparison between these new temperatures and those from the original
\citet{ikebe02} parent sample is shown plotted in Fig.~\ref{fig:compare_kT}.
There is slightly better agreement between our new temperatures and those of 
\citeauthor{ikebe02} amongst the hottest clusters than was obtained using 
the measurements from \citetalias{san06}, although the difference between
this plot and figure~2 from \citetalias{san06} is fairly small\footnote{NB
the cool-core status of the points in figure~2 of \citet{san06} was
incorrectly labelled.} The \citet{ikebe02} mean temperatures were based on
a two-component spectral fit to the full cluster emission, which would
likely result in lower \Tbar\ values for the hottest clusters, as pointed
out in \citetalias{san06}. An additional factor is a possible bias in
temperature estimates for hotter clusters ($\ga$4-5 \keV) resulting from
errors in the \Chandra\ response matrix \citep[as described in][and
references therein]{sun09}, which could lead to overestimates of
\Tbar. Our final \Tbar\ values are listed in Table~\ref{tab:main}, together 
with the mean metallicity measured in the same aperture (see
Section~\ref{ssec:Z(r)}).

Although the \MT\ relation of \citet{vikhlinin06} comprises clusters hotter
than $\sim$2.3 keV, we none the less have verified that the masses of our
cooler galaxy groups (O'Sullivan et al. in prep.) are consistent with
it. Furthermore, we note that the recent \Chandra\ study of 40 galaxy
groups by \citet{sun09} finds only weak evidence (at the 1.5$\sigma$ level)
for a steepening of the \MT\ relation in groups compared to clusters. This
therefore justifies our inclusion of a mixture of clusters and groups in
the 7 systems used to calibrate our modification of the
\citeauthor{vikhlinin06} \MT\ relation for temperatures measured in the
aperture 0.15--0.2\rfiveh.

\begin{table*}
\begin{center}
\begin{tabular}{lccccccccc}
  \hline\\[-2ex]
Name & ObsID$^a$ & Detector$^b$ & Redshift & Mean $kT^c$ & \rfiveh\ & Mean metallicity$^c$ & Cool-core status & S$_{0.1\rfiveh}^d$ & Index$^d$ \\ \\[-2ex]
 & & & & (keV) & (kpc) & (Solar) & & (\keV\cmsq) & \\ \\[-2ex]
  \hline \\[-2ex]
NGC 5044 & 798 & S & 0.008 & 1.17$_{-0.05}^{+0.04}$ & 512$_{-10}^{+10}$ & 0.37$_{-0.06}^{+0.09}$ & CC & 50$_{-17}^{+27}$ & 0.71$_{-0.09}^{+0.39}$ \\
  Abell 262 & 2215 & S & 0.016 & 2.08$_{-0.09}^{+0.11}$ & 692$_{-15}^{+18}$ & 0.38$_{-0.07}^{+0.10}$ & CC & 126$_{-20}^{+48}$ & 0.91$_{-0.13}^{+0.30}$ \\
  Abell 1060 & 2220 & I & 0.012 & 2.92$_{-0.11}^{+0.11}$ & 829$_{-16}^{+16}$ & 0.50$_{-0.07}^{+0.07}$ & non-CC & 191$_{-7}^{+9}$ & 0.44$_{-0.07}^{+0.03}$ \\
  Abell 4038 & 4992 & I & 0.030 & 3.04$_{-0.09}^{+0.07}$ & 840$_{-13}^{+10}$ & 0.60$_{-0.07}^{+0.07}$ & non-CC & 136$_{-4}^{+8}$ & 0.55$_{-0.14}^{+0.08}$ \\
  Abell 1367 & 514 & S & 0.022 & 3.22$_{-0.18}^{+0.18}$ & 869$_{-26}^{+25}$ & 0.34$_{-0.09}^{+0.10}$ & non-CC & 275$_{-31}^{+10}$ & 0.28$_{-0.31}^{+0.10}$ \\
  Abell 2147 & 3211 & I & 0.035 & 3.69$_{-0.18}^{+0.18}$ & 928$_{-23}^{+23}$ & 0.28$_{-0.10}^{+0.11}$ & non-CC & 281$_{-59}^{+32}$ & 0.43$_{-0.16}^{+0.27}$ \\
  2A 0335+096 & 919 & S & 0.035 & 4.09$_{-0.13}^{+0.13}$ & 980$_{-16}^{+16}$ & 0.79$_{-0.11}^{+0.11}$ & CC & 99$_{-20}^{+20}$ & 1.12$_{-0.16}^{+0.22}$ \\
  Abell 2199 & 497 & S & 0.030 & 4.50$_{-0.24}^{+0.20}$ & 1033$_{-29}^{+23}$ & 0.79$_{-0.14}^{+0.14}$ & CC & 165$_{-33}^{+17}$ & 0.76$_{-0.06}^{+0.09}$ \\
  Abell 496 & 4976 & S & 0.033 & 4.80$_{-0.14}^{+0.15}$ & 1067$_{-16}^{+16}$ & 0.66$_{-0.08}^{+0.09}$ & CC & 162$_{-34}^{+55}$ & 0.93$_{-0.07}^{+0.32}$ \\
  Abell 1795 & 5289 & I & 0.062 & 5.62$_{-0.35}^{+0.36}$ & 1144$_{-38}^{+37}$ & 0.26$_{-0.10}^{+0.10}$ & CC & 189$_{-35}^{+21}$ & 0.99$_{-0.08}^{+0.33}$ \\
  Abell 3571 & 4203 & S & 0.039 & 6.41$_{-0.23}^{+0.23}$ & 1239$_{-23}^{+23}$ & 0.75$_{-0.11}^{+0.11}$ & non-CC & 279$_{-7}^{+9}$ & 0.44$_{-0.04}^{+0.04}$ \\
  Abell 2256 & 1386 & I & 0.058 & 6.52$_{-0.36}^{+0.39}$ & 1239$_{-36}^{+38}$ & 0.98$_{-0.21}^{+0.22}$ & non-CC & 344$_{-158}^{+24}$ & 0.47$_{-0.16}^{+0.47}$ \\
  Abell 85 & 904 & I & 0.059 & 6.64$_{-0.20}^{+0.20}$ & 1251$_{-19}^{+19}$ & 0.56$_{-0.07}^{+0.07}$ & CC & 193$_{-20}^{+16}$ & 0.90$_{-0.09}^{+0.14}$ \\
  Abell 3558 & 1646 & S & 0.048 & 7.17$_{-0.46}^{+0.49}$ & 1309$_{-44}^{+46}$ & 1.00$_{-0.20}^{+0.21}$ & non-CC & 304$_{-53}^{+28}$ & 0.62$_{-0.11}^{+0.30}$ \\
  Abell 3667 & 889 & I & 0.056 & 7.60$_{-0.37}^{+0.38}$ & 1345$_{-35}^{+34}$ & 0.51$_{-0.10}^{+0.10}$ & non-CC & 407$_{-67}^{+46}$ & 0.57$_{-0.09}^{+0.17}$ \\
  Abell 478 & 1669 & S & 0.088 & 8.23$_{-0.26}^{+0.26}$ & 1381$_{-23}^{+23}$ & 0.50$_{-0.07}^{+0.07}$ & CC & 190$_{-8}^{+18}$ & 1.03$_{-0.06}^{+0.21}$ \\
  Abell 3266 & 899 & I & 0.055 & 8.38$_{-0.43}^{+0.67}$ & 1417$_{-38}^{+58}$ & 0.39$_{-0.11}^{+0.11}$ & non-CC & 528$_{-36}^{+19}$ & 0.48$_{-0.09}^{+0.09}$ \\
  Abell 2029 & 4977 & S & 0.077 & 8.96$_{-0.30}^{+0.30}$ & 1452$_{-25}^{+25}$ & 0.60$_{-0.07}^{+0.07}$ & CC & 257$_{-49}^{+20}$ & 0.90$_{-0.08}^{+0.21}$ \\
  Abell 401 & 2309 & I & 0.074 & 9.16$_{-1.06}^{+1.41}$ & 1471$_{-92}^{+114}$ & 0.26$_{-0.23}^{+0.22}$ & non-CC & 427$_{-56}^{+55}$ & 0.45$_{-0.11}^{+0.15}$ \\
  Abell 2142 & 5005 & I & 0.091 & 9.50$_{-0.42}^{+0.43}$ & 1487$_{-35}^{+34}$ & 0.44$_{-0.07}^{+0.07}$ & non-CC & 295$_{-18}^{+19}$ & 0.94$_{-0.17}^{+0.06}$ \\ \\[-2ex]
   \hline
\end{tabular}
\caption{Key properties of the sample, listed in order of increasing
temperature. $^a$\Chandra\ observation identifier. $^b$ACIS detector. 
$^c$Measured between 0.15 and 0.2\rfiveh\ (see Section~\ref{sec:mean_kT}). 
$^d$Parameters of the power-law fit to the entropy profile (see 
Section~\ref{ssec:S(r)_fit}). Errors are 1$\sigma$.}
\label{tab:main}
\end{center}
\end{table*}

\subsection{Spectral profiles and deprojection analysis}
\label{ssec:spec_prof}
The results which follow are based on the deprojection analysis method
described in \citetalias{san06}, to derive three-dimensional gas
temperature and density profiles. To stablize the fitting, the Galactic
absorbing column and gas metallicity were fixed at values obtained by
fitting each annulus separately prior to the deprojection. Consequently,
the gas metallicity results presented in Section~\ref{ssec:Z(r)} are
projected quantities.  For some clusters, it was necessary to freeze the
absorbing column at the galactic HI value, since unfeasibly low values were
otherwise obtained in many of the annular bins: full details can be found
in \citetalias{san06}. Similarly, in a number of cases the deprojected
temperature had to be fixed at its projected value, to produce a stable
fit, exactly as was done in \citetalias{san06}.

\section{Results}

\subsection{Gas density}
\label{ssec:rho_gas(r)}
\begin{figure}
\centering
\includegraphics[width=8.4cm]{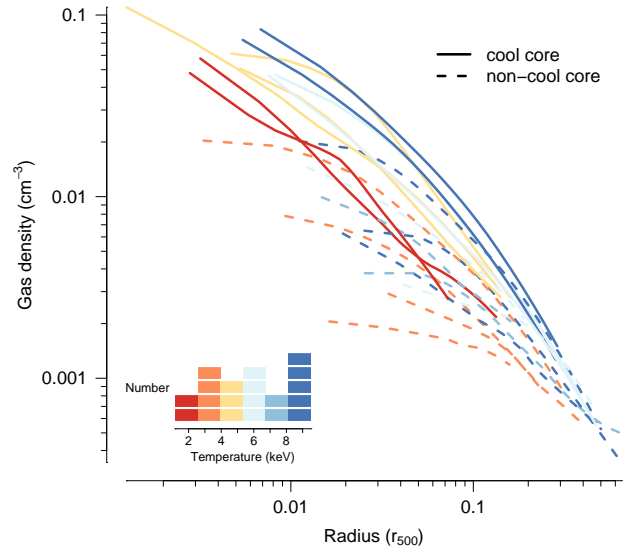}
\caption{ \label{fig:rho.gas_vs_r500}
Gas density profiles for each cluster, scaled to \rfiveh\ and coloured
according to the mean cluster temperature, depicted by the inset
histogram. Each curve represents a locally weighted fit to the data points,
to suppress small-scale fluctuations (see the text for details).  }
\end{figure}

Fig.~\ref{fig:rho.gas_vs_r500} shows the gas density as a function of
scaled radius for the sample, colour-coded by mean temperature and split
according to cool-core status. Following \citet{san06}, in order to clarify
the underlying trend in each profile, the raw data points have been fitted
with a locally weighted regression in log-log space, using the task
`\textsc{loess}' in version 2.7 of the \Rproject\ statistical software
environment package\footnote{http://www.r-project.org} \citep{Rcite}. It is
immediately apparent from Fig.~\ref{fig:rho.gas_vs_r500} that the profiles
do not scale self-similarly, which would imply constant density at a given
fraction of \rfiveh. There is a wide dispersion in the core, with a factor
of $\sim$30 range in \rhogas\ at 0.01\rfiveh, dropping to a factor of
$\sim$5 spread at 0.1\rfiveh. At larger radius \citepalias[beyond the peak
in the temperature profile at $\sim$0.15\rfiveh;][]{san06} there is a clear
convergence in the profiles, despite the diminishing data coverage due to
the limits of the \Chandra\ field of view.

It is clear from Fig.~\ref{fig:rho.gas_vs_r500} that the cool-core
(hereafter CC) clusters have systematically denser and more cuspy cores,
consistent with the decline in gas temperature \citepalias[scaling with
r$^{\sim 0.4}$;][]{san06} that must be counteracted by a rising density in
order to maintain pressure equilibrium. The near power-law shape of the
density profiles in the CC clusters is also consistent with a cooling
dominated regime, as indicated by the simulations of \citet{ettori08}, who
find $\rhogas \propto r^{-1.2}$ for evolved cool-core clusters (after
10\,Gyr). Within each cool-core category, there is also a systematic trend
towards lower ICM densities in cooler systems. This may reflect gas
depletion due to cooling out of the hot phase, or could be caused by
non-gravitational heating expelling material out of the core. These
findings are consistent with the recent study of \citet{croston08}, based 
on a representative sample of 31 clusters analysed with \XMM.

\subsection{Gas entropy}
\label{ssec:egas(r)}
\begin{figure}
\centering
\includegraphics[width=8.4cm]{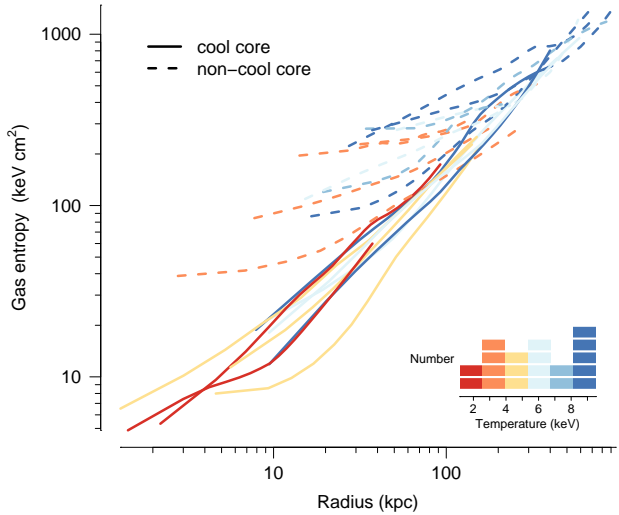}
\caption{ \label{fig:S_vs_kpc}
Gas entropy profiles as a function of physical radius, colour-coded by
cluster mean temperature. Each curve represents a locally weighted fit to
the data points, to suppress small-scale fluctuations (see the text for
details).
}
\end{figure}

The entropy of the gas is a key parameter, which provides a measure of the
thermodynamic state of the ICM and is conserved in any adiabatic process
\citep[see][for example]{bower97,tozzi01,voit02}.  We define entropy as
\hbox{$S=kT/\rhogas^{2/3}$} \citep[e.g.][]{pon99}, which implies $S\propto
kT$ for self-similar clusters. Within a virialized halo, the gas entropy is
initially set by shock heating, which leads to a radial variation of the
form $S\propto r^{1.1}$ for a simple spherical collapse
\citep{tozzi01}. Despite radiative cooling in the cores of 
clusters lowering the entropy, its power-law variation with radius is
nevertheless largely preserved, with approximately the same logarithmic
slope as in the outskirts \citep[e.g.][]{ettori08}, leading to the
expectation of a near-proportionality between entropy and physical
(i.e. unscaled) radius, regardless of halo mass. This can be seen in
Fig.~\ref{fig:S_vs_kpc} which shows gas entropy profiles as a function of
radius in kpc for the sample, separated by cool-core status and
colour-coded by mean temperature.

The distinction between CC and non-CC clusters is very clear from
Fig.~\ref{fig:S_vs_kpc}, with the latter having significantly higher
entropy in the core and the divergence between the two types occurring
within $\sim$40\kpc, at an entropy level of $\sim$80\keV\cmsq. The CC
entropy profiles are tightly grouped and show no obvious sign of the
transition between the shock heating regime in the outskirts and the
cooling-dominated core. In contrast, the non-CC profiles show a much larger
scatter in normalization, which may reflect the diversity of heating
processes affecting them. Alternatively, it could be that cooling acts to
regulate the entropy of CC clusters so as to suppress the scatter between
them, which may also account for the apparent universality seen in the
cooling time profiles of the sample \citepalias{san06}.

\subsubsection{Entropy profile fitting}
\label{ssec:S(r)_fit}
In order to characterize the form of the entropy profiles plotted in
Fig.~\ref{fig:S_vs_kpc}, we fitted power-laws to each cluster profile to
quantify the logarithmic slope of the relationship. However, it can be seen
from Fig.~\ref{fig:S_vs_kpc} that at small radii the profiles tend to
flatten and therefore deviate from a simple power law form \citep[see
also][for example]{donahue06}. Nevertheless, a power law provides a simple
and reasonably effective description of the majority of the profile in all
cases which enables the `flatness' of the curves to be characterized. We
have additionally employed a quantile regression technique to perform the
fitting, in order to provide resistance to any such outliers. This form of
regression minimizes the sum of absolute residuals, rather than the sum of
squared residuals, and thus is analogous to determining the median as an
estimator of the mean \citep{koenker05}. The algorithm used is the
`\textsc{rq}' function from the \textsc{quantreg} package in
\Rproject\footnote{See the tutorial at 
http://www.astrostatistics.psu.edu/su07/R/reg.html} and was executed as an
unweighted linear fit in log-log space. Table~\ref{tab:main} lists the
best-fit normalization (at 0.1\rfiveh) and power-law index
(i.e. logarithmic slope) for each cluster, together with the corresponding
1$\sigma$ errors.

\begin{figure}
\centering
\includegraphics[width=8.4cm]{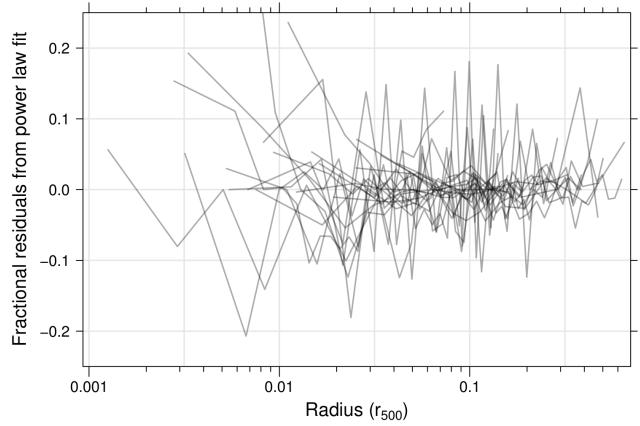}
\caption{ \label{fig:S_fit_resid}
Residuals from the power law fits to the entropy profiles, normalized by the
predicted value and plotted against scaled radius for each cluster. A
single value is omitted (0.76 at 0.0047\rfiveh, for the innermost point of
2A~0335+096) to optimize the y-axis scale range for viewing the data.}
\end{figure}

To assess the suitability of a power law for describing the entropy
profiles, we show in Fig.~\ref{fig:S_fit_resid} the residuals from the
best-fit model as a function of scaled radius, for each cluster. We have
normalized the residuals by the predicted model values rather than the
measurement errors on the entropy, since the fit was performed using
unweighted quantile regression rather than by minimizing \chisq. This
approach allows for the fact that real clusters exhibit intrinsic
deviations from simple radial models in excess of the statistical scatter
associated with measurement errors. It can be seen that in general the
power law fits do a reasonable job of describing the data, with the
majority of residuals contained within $\sim$5 per cent of the best
fit. However, while there is no obvious sign of any strong systematic
trends with radius, there is some indication of an excess entropy above the
model at very small radii ($\la$0.03\rfiveh) for a few of the clusters,
consistent with the flattening in the entropy profiles of nine cool core
clusters found by \citet{donahue06}. Nevertheless, we restrict our use of
the power law fits to the entropy profile to a larger radius (0.1\rfiveh),
which is not affected by any such systematic departure from a power law
form.

\begin{figure}
\centering
\includegraphics[width=8cm]{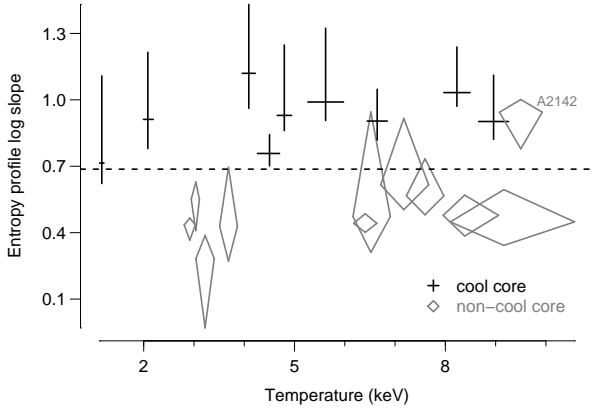}
\caption{ \label{fig:S-index_vs_kT}
Best fit logarithmic slope of the entropy profile as a function of mean
system temperature. The dashed line marks the cross-over point between the
two Gaussian distributions of values from Section~\ref{sec:bimodality} and
the outlier Abell~2142 is labelled (see text for details).}
\end{figure}

There is no evidence of any systematic variation in the logarithmic slope
of the entropy profile with mean temperature across the sample, as plotted
in Fig.~\ref{fig:S-index_vs_kT}. However CC and non-CC clusters are clearly
segregated in the plot, occupying higher and lower values of the entropy
index, respectively. The only exception is Abell~2142 -- the archetypal
`cold front' cluster \citep{markevitch00} -- which was classified as a
non-CC cluster in \citetalias{san06}, as its cool core ratio was found to
be only marginally significant
\citepalias[$\sim2\sigma$;][]{san06}. Notwithstanding the cool core status
of the clusters, it is reasonable to ask if there is any evidence for
bimodality in the distribution of logarithmic slope values in
Fig.~\ref{fig:S-index_vs_kT}, and we address this specific issue in
Section~\ref{sec:bimodality}.

\subsubsection{Entropy scaling}
\label{ssec:S_scaling}
The effectiveness of entropy as a probe of non-gravitational physics in the
ICM is best exploited by studying its variation with mean temperature,
which is expected to be linear in the case of simple
self-similarity. However, the strong dependence of entropy on radius,
raises the issue of at what point to measure entropy in order to consider
its variation from cluster to cluster. The issue is further complicated by
the fact that the gas entropy within a cluster bridges two distinct
regimes: the cooling-dominated core and the outskirts, where the effects of
shock-heating from infall prevail. The first measurements of entropy
scaling were made at a fiducial radius of 0.1\rtwoh\
\citep{pon99,llo00} in order to sample the region between these two
regimes. However, \citet{pratt06} find consistent results for the scaling
with temperature of entropy when measured a range of radii at fractions of
\rtwoh\ of 0.1, 0.2, 0.3 and 0.5 in the range $b$\,=\,0.5--0.6 for
$S\propto\Tbar^b$.

\begin{figure}
\centering
\includegraphics[width=8cm]{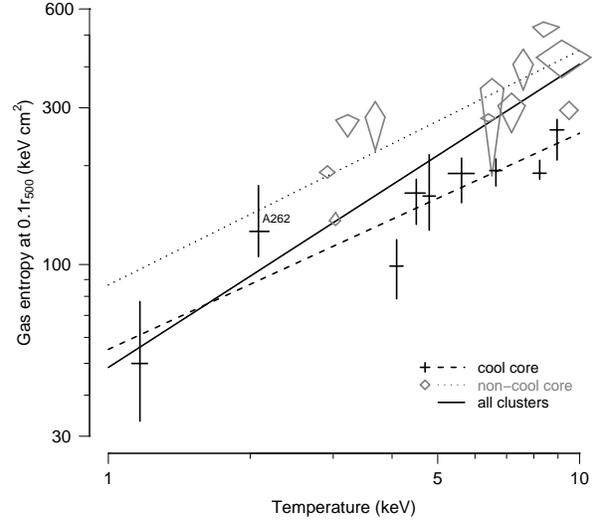}
\caption{ \label{fig:egas_vs_kT}
Entropy at 0.1\rfiveh\ as a function of mean temperature, from a power-law
fit to the entropy profile of each cluster. The lines are the
best-fit power-law, with logarithmic slopes of $0.66\pm0.10$, $0.71\pm0.21$ 
and $0.92\pm0.12$ for the CC, non-CC and combined clusters, respectively.}
\end{figure}

Taking the power-law fits to the entropy profiles performed above we plot
in Fig.~\ref{fig:egas_vs_kT} entropy vs. \Tbar\ for the sample using the
normalization value, corresponding to a fiducial radius of 0.1\rfiveh. Also
shown are power-law fits to the CC and non-CC clusters separately,
performed in log-log space using the BCES weighted orthogonal regression
method of \citet{akritas96}. The logarithmic slopes of these lines
correspond to an entropy scaling of the form $S\propto\Tbar^b$ with
$b=0.66\pm0.10$ and $0.71\pm0.21$ for the CC and non-CC
clusters, respectively. These results are in good agreement with the value
of $0.65\pm0.05$ similarly determined by \citet{pon03}, at a fiducial
radius of 0.1\rtwoh, as well as with the values found by \citet{pratt06} at
a series of radii and using a number of regression methods. 

The CC and non-CC points are generally well separated with respect to their
corresponding best-fit power laws in Fig.~\ref{fig:egas_vs_kT}. However, in
the case of the CC cluster Abell~262, the (albeit large) error bar overlaps
significantly with the \emph{non-}CC best-fit line. While this poor cluster does
have a strong negative central temperature gradient, it is also known to
possess a number of cavity and ripple features coincident with low
frequency radio emission \citep{blanton04}. It is therefore possible that
AGN activity in the core might have boosted the entropy at 0.1\rfiveh\ to
shift this cluster towards the non-CC regression line. Nevertheless, a
number of the other CC clusters also host ghost cavities (see table~2 in
\citetalias{san06}), which evidently have not significantly biased their
location in Fig.~\ref{fig:egas_vs_kT}.

While the separate fits to the CC and non-CC clusters in
Fig.~\ref{fig:egas_vs_kT} yields results consistent with a \emph{non-}self
similar entropy scaling, the two sets of points are clearly offset from
each other.  It is interesting to note that a BCES orthogonal regression
fit to all the 20 clusters combined yields a much steeper logarithmic slope
of $0.92\pm0.12$, which \emph{is} consistent with self-similar scaling. 
To check this result, we repeated the individual power-law fits using a
pivot-point of 0.15\rfiveh, and performed the same regression of the
normalization versus mean temperature to obtain the following logarithmic
slopes:
\begin{description}
\item $1.06\pm0.16$ (all clusters combined)
\item $0.65\pm0.10$ (CC)
\item $0.65\pm0.26$ (non-CC).
\end{description}
Thus it is possible that similarity breaking can occur in the separate
CC/non-CC cluster sub-populations in such a way as to produce fully
self-similar scaling in an analysis which combines the two types.
On the other hand, if we exclude the coolest system (NGC~5044) from the
fit, the resulting best-fit slope for the whole sample ($0.89\pm0.22$) 
is consistent at the $1\sigma$ level with the values obtained for 
the separate CC and non-CC sub-samples.

\begin{figure}
\centering
\includegraphics[width=8.4cm]{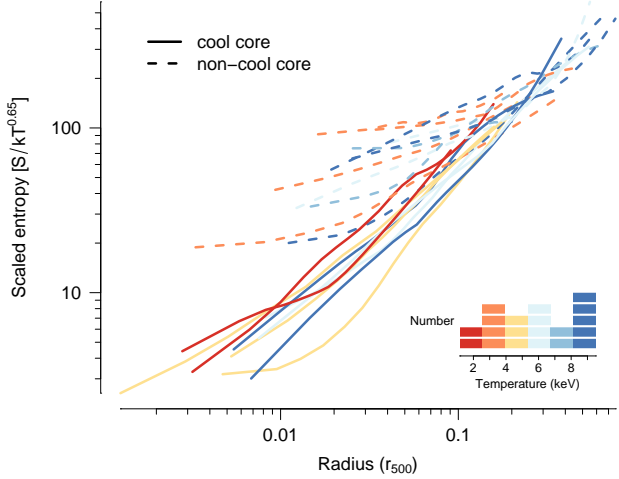}
\caption{ \label{fig:sc065S_vs_r500}
Entropy profiles, scaled with the \citet{pon03} empirical factor of
\Tbar$^{0.65}$ vs. scaled radius, colour-coded by
cluster mean temperature. Each curve represents a locally weighted fit to
the data points, to suppress small-scale fluctuations (see the text for
details).  }
\end{figure}

Given the broad agreement in the dependence on temperature of the entropy
scaling, we show in Fig.~\ref{fig:sc065S_vs_r500} entropy profiles as a
function of $r$\,/\,\rfiveh\ normalized by the factor $T^{0.65}$ from
\citet{pon03}. It can be seen that this empirical scaling brings the curves
into fairly close alignment. There is, however, some indication of a
systematic dispersion in the profiles, indicating that gas entropy may vary
less strongly with cluster mean temperature than $T^{0.65}$, as also
suggested by the results of \citet{pratt06} when measuring entropy at
0.1\rtwoh.  In order to explore the entropy scaling across a range of
radii, rather than at at particular spot value, we modelled the variation
of entropy in terms of both scaled radius
\emph{and} mean temperature using the following expression:
\begin{equation}
\label{eqn:S(r)_fit}
S = S^\prime \times \left( \frac{r}{0.1\rfiveh} \right)^a \times 
 \left( \frac{\Tbar}{5\keV} \right)^b .
\end{equation}
This parametrization has the advantage of allowing the scaling of entropy
with both radius and temperature to be determined simultaneously. The
results of performing an unweighted quantile regression fit in log-log
space of Equation~\ref{eqn:S(r)_fit} to the entire sample are summarised in
Table~\ref{tab:S(r)_fit}. Only data in the range $0.02 < r/\rfiveh\ < 0.2$
were included, to ensure completeness in the radial coverage. 

\begin{table}
\begin{center}
\begin{tabular}{lccc}
  \hline \\[-2ex]
Sample & $S^\prime$\,(\keV\cmsq) & $a$ & $b$ \\
  \hline \\[-2ex]
All & 214$_{-11}^{+10}$ & 0.95$_{-0.08}^{+0.10}$ & 0.56$_{-0.08}^{+0.11}$ \\
  CC & 161$_{-10}^{+7}$ & 1.05$_{-0.15}^{+0.07}$ & 0.49$_{-0.13}^{+0.07}$ \\
  NCC & 259$_{-4}^{+6}$ & 0.51$_{-0.06}^{+0.09}$ & 0.52$_{-0.07}^{+0.08}$ \\ \\[-2ex]
   \hline
\end{tabular}
\caption{Results from fitting Equation~\ref{eqn:S(r)_fit} to the 
entire sample, as well as separately to the combined cool-core and 
non-cool core clusters. Errors are 1$\sigma$; see text for details.}
\label{tab:S(r)_fit}
\end{center}
\end{table}

The resulting values of $a$ agree well with the results from the individual
power law fits to each cluster described previously, in
Section~\ref{ssec:S(r)_fit}. It can also be seen that the fit to the entire
sample yields a dependency on mean temperature such that $S\propto\Tbar^b$,
with $b=0.56_{-0.08}^{+0.11}$ slightly lower but still in agreement with
the scaling of $\Tbar^{0.65\pm0.05}$ found by \citet{pon03}, which was
evaluated at a radius of 0.1\rtwoh\ --- roughly 0.15\rfiveh\
\citep{san03b}. A fit to the combined CC and non-CC clusters separately yields
values of $b=0.49_{-0.13}^{+0.07}$ and $b=0.52_{-0.07}^{+0.08}$,
respectively, demonstrating that this weaker dependence of entropy on
system temperature persists. This result is strongly inconsistent with a
self-similar scaling of $S\propto\Tbar$ (i.e. $b=1$) demonstrating the
impact of non-gravitational physics in influencing the hot gas
\citep[see][for example, for further discussion]{pon03}, but extends
beyond the finding of \citeauthor{pon03} in that it applies across a range
of radii spanning an order of magnitude, as opposed to being determined at
only a fixed spot value. Furthermore, the fact that this result is
essentially unchanged when fitting the CC and non-CC clusters separately
(see Table~\ref{tab:S(r)_fit}), demonstrates the universality of this
modified entropy scaling, at least in the range $0.02<r/\rfiveh<0.2$.

However, the difference between CC and non-CC clusters is very clear when
considering the variation of entropy with radius, $S\propto r^a$. Here the
corresponding logarithmic slopes are $a=1.05_{-0.15}^{+0.07}$ and
$0.51_{-0.06}^{+0.09}$, respectively, compared to $0.95_{-0.08}^{+0.10}$ for
the entire sample. The flatness of the best-fit non-CC entropy profile
compared to that for the CC clusters is further underscored by the
difference in normalization at the fiducial `pivot point' in
Equation~\ref{eqn:S(r)_fit} (at 0.1\rfiveh\ for $\Tbar=5\keV$), yielding
values of $S^\prime=259_{-4}^{+6}$ \keV\cmsq\ compared to $161_{-10}^{+7}$
\keV\cmsq, respectively. The results from the combined fit of
Equation~\ref{eqn:S(r)_fit} are in good agreement with the results from the
power-law fits to the individual cluster entropy profiles. They are also
consistent with the logarithmic slope of 0.95$\pm$0.02 \citep{piffaretti05}
and 1.08$\pm$0.04 \citep{pratt06} from two separate \XMM\ analyses of
cool-core clusters, as well as the values of slopes in the range 1--1.3
from the \Chandra\ analysis of 9 CC clusters by
\citet{donahue06}.

\subsection{Metallicity of the ICM}
\label{ssec:Z(r)}
\begin{figure}
\centering
\includegraphics[width=8cm]{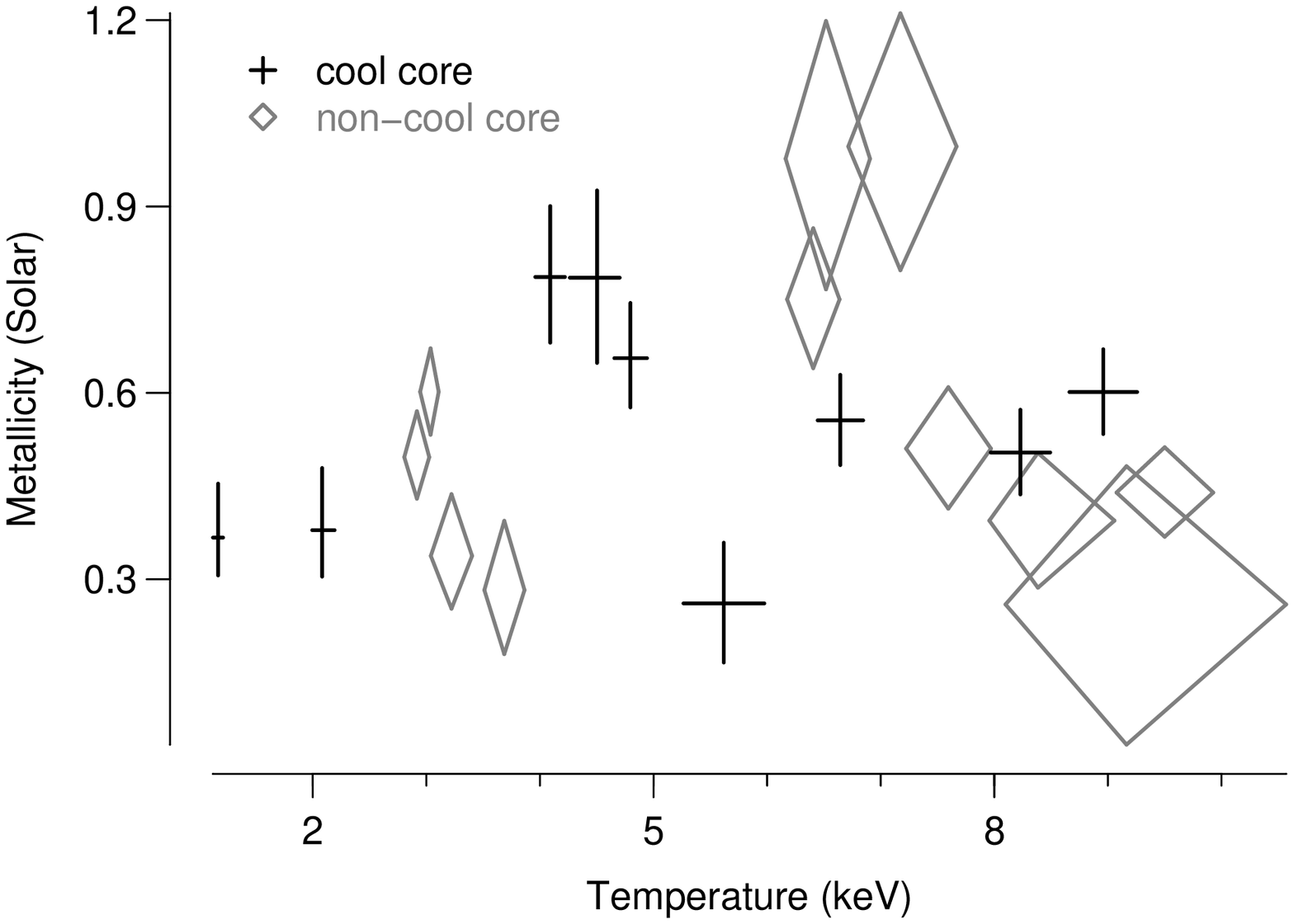}
\caption{ \label{fig:Z_vs_kT}
Gas metallicity (\citealt{grevesse98} abundances) versus mean temperature,
both measured in the range $0.15 \le r/\rfiveh \le 0.2$.}
\end{figure}

The metallicity of the ICM is an important tracer of galaxy feedback via
supernova-driven winds, which eject both metals and energy into the hot gas 
\citep[e.g.][]{strickland00}. Fig.~\ref{fig:Z_vs_kT} shows the mean 
metallicity against mean temperature for the sample (see
Table~\ref{tab:main}), both evaluated in the range $0.15\le r/\rfiveh\le 0.2$,
identified according to cool-core status. There is no indication of any
trend in the data or of any systematic difference between CC and non-CC
clusters; the mean value across the entire sample is 0.55 Solar, with a
standard deviation of 0.22.

To explore radial trends we use the metallicity values determined in the
projected annular spectral fitting, which were taken as fixed inputs in the
deprojection analysis as described in Section~\ref{ssec:spec_prof} and
\citetalias{san06}. The fact that these values are not deprojected
will tend to smooth out gradients slightly. However, \citet{rasmussen07}
demonstrate that this is a small effect and that deprojected metallicity
data typically suffer from large instabilities which add significantly to
the noise, particularly given the poorer constraints on this parameter
compared to temperature or density.

\begin{figure}
\centering
\includegraphics[width=8.4cm]{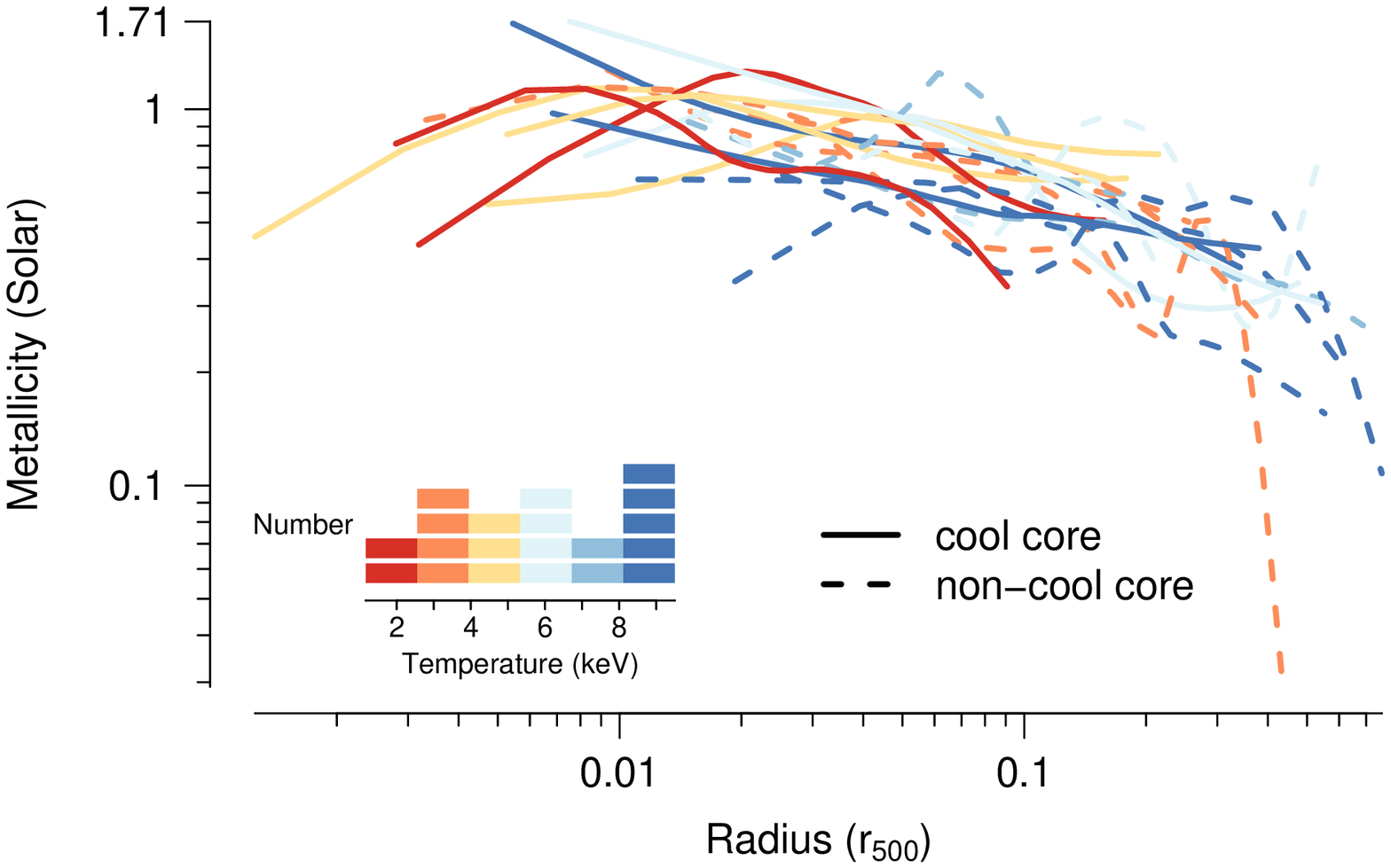}
\caption{ \label{fig:Z(r)}
Projected gas metalllicity profiles (using \citealt{grevesse98} abundances)
for each cluster, scaled to \rfiveh\ and coloured according to the mean
cluster temperature, depicted by the inset histogram. Each curve represents
a locally weighted fit to the data points, to suppress small-scale
fluctuations (see the text for details).  }
\end{figure}

Fig.~\ref{fig:Z(r)} shows the individual projected metallicity profiles as
a function of scaled radius for the sample, split by cool-core status and
colour-coded by mean temperature. The curves represent a locally-weighted
fit to the data, using the \textsc{loess} function described in
Section~\ref{ssec:rho_gas(r)} and it can clearly be seen that the
metallicity declines with radius beyond $\sim$0.01--0.02\rfiveh\ in almost
every case, for both CC and non-CC clusters. Within this radius the
dispersion in metallicity increases noticeably; some clusters show a
strongly peaked metallicity profile (in particular Abell~85 and
Abell~2029), whereas others have metal-deficient central cores. More
specifically, the three most peaked profiles are all hot clusters, whereas
the coolest clusters show sharp central declines. It is not clear what is
responsible for this strong divergence in behaviour, but it occurs on a
scale of roughly 0.02\rfiveh\ ($\sim$20--30\,kpc), which is comparable to
the size of the central galaxy.  This also coincides with the region where
the gas density and entropy profiles also become widely dispersed
(Figs.~\ref{fig:rho.gas_vs_r500} and \ref{fig:sc065S_vs_r500}).

Regardless of any difference between CC and non-CC clusters at small radii,
it is clear that the point where the metallicity begins to decline strongly
towards the cluster outskirts lies well within the typical cooling radius
of $\sim$0.15\rfiveh\ \citepalias{san06}. The fact that the metallicity
profiles of all the clusters are quite similar beyond this point suggests
that the enrichment of the ICM is insensitive to those factors responsible
for determining cool-core status. The finding that metallicity declines
strongly with radius in all the clusters is apparently at odds with the
study of \citet{degrandi01}, who concluded that the abundance profiles for
the 8 non-CC clusters in their \BeppoSAX\ sample were consistent with being
constant. However, examination of figure~2 from \citet{degrandi01} shows
that, with the exception of only a few outliers, there is an indication of
a general decline in metallicity with increasing radius in their non-CC
clusters. Furthermore, we note that \citet{baldi07} also found no
difference between CC and non-CC gas metallicity profiles, outside
$\sim$0.1r$_{\mathrm{180}}$ in their analysis of 12 hot clusters observed
with \Chandra.


\begin{figure}
\centering
\includegraphics[width=8cm]{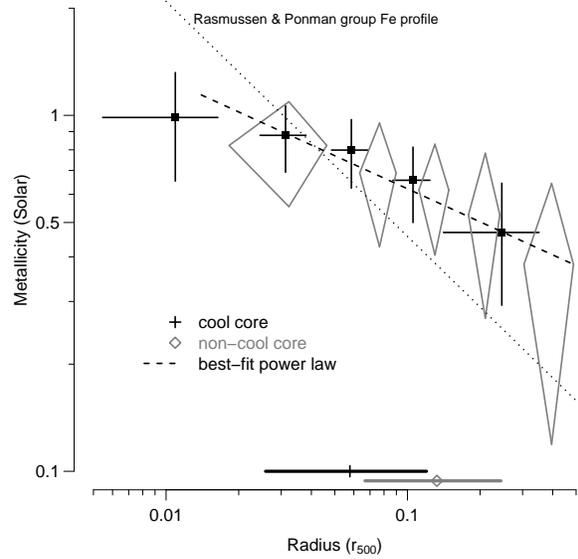}
\caption{ \label{fig:binned_Z(r)}
Average gas metallicity profiles for cool-core and non-cool core clusters.
Each bin represents the mean of a roughly equal number of points and the
error bars show the standard deviation in both directions. The dashed line
is the best-fit power-law to the unbinned data beyond 0.014$r/\rfiveh$ and
the dotted line is a fit to the iron abundances in 15 galaxy groups from
\citet[see text for details]{rasmussen07}. The points and horizontal lines 
indicate the medians and interquartile ranges of the CC and non-CC raw data
points.}
\end{figure}

To highlight the trend in metallicity with radius we show the mean CC and
non-CC profiles in Fig.~\ref{fig:binned_Z(r)}, grouped to a total of 5 bins
in each case, with error bars indicating the standard deviation within each
bin. Apart from a slightly larger dispersion in the non-CC bins, there is
very little difference between the two profiles and no indication in either
case of any flattening off at large radii. Interestingly, the \BeppoSAX\
average CC profile of \citet{degrandi04} shows a sharp levelling off in
metallicity at around 0.4 Solar \citep[abundances]{grevesse98} at
0.2\rtwoh, which corresponds to roughly 0.3\rfiveh\ -- close to the limit
of our \Chandra\ data but otherwise consistent with the outer bin of
Fig.~\ref{fig:binned_Z(r)}, so we cannot rule out a flattening beyond this
point. 

However, in contrast, the \citet{degrandi04} non-CC profile is both flatter
and lower in normalization (albeit only at $\sim$2$\sigma$) and thus rather
different from our own. Of their 10 non-CC clusters, only three are present
in our sample (A1367, A2256 and A3266; they classifiy A2142 and A3571 as CC
clusters), and their profiles for these systems appear to be reasonably
consistent with our own, within the region covered by
\Chandra. The differences between CC and non-CC clusters seen by
\citet{degrandi04} appear to originate in the somewhat low and fairly flat
abundance profiles of several non-CC clusters in their sample, whereas the
non-CC clusters in our sample don't appear to include such
cases. Notwithstanding this difference in samples, we note that
\citet{baldi07} measure a similar radial decline to us, with essentially no
difference between CC and non-CC clusters, albeit with the exception of
their central bin. As mentioned above, cluster metallicity profiles appear
to exhibit greater dispersion on the scale of the central galaxy, so some
difference from sample to sample is to be expected in this region.

The data in Fig.~\ref{fig:binned_Z(r)} appear to be reasonably consistent
with a power law, so we have fitted such a function in log-log space, using
the quantile regression method outlined in Section~\ref{ssec:S(r)_fit}. In
order to exclude the core region where the profile flattens and the profile
diverge substantially ($r\sim0.014\rfiveh$; Fig.~\ref{fig:Z(r)}), we
performed separate fits inside and outside this radius, combining both the
CC and non-CC data. The results for the inner region ($r<0.014\rfiveh$) are
\begin{displaymath}
\log_{10}(Z/Z_{\sun}) = 0.39_{-0.22}^{+0.03} \log_{10}(r/\rfiveh) + 0.82_{-0.4}^{+0.06},
\end{displaymath}
while those for the outer region ($r\geq0.014\rfiveh$) are
\begin{displaymath}
\log_{10}(Z/Z_{\sun}) = -0.31_{-0.05}^{+0.03} \log_{10}(r/\rfiveh) -0.51_{-0.07}^{+0.03},
\end{displaymath}
By comparison, the metallicity profiles of galaxy groups appear to be
somewhat steeper in their decline at large radii. The recent analysis of 15
groups with \Chandra\ data by \citet{rasmussen07}-- also using
\citet{grevesse98} abundances-- found logarithmic slopes of -$0.66\pm0.05$ and
-0.44 for the combined radial profiles of iron and silicon, respectively,
across their sample, compared to our slope of -$0.31_{-0.05}^{+0.03}$ for
the mean metallicity (see Fig.~\ref{fig:binned_Z(r)}). At a radius of
\rfiveh, \citeauthor{rasmussen07} estimate an average iron abundance in groups to be
$\sim$0.1, whereas the extrapolation of the above best fit implies a mean
metallicity of $Z\sim0.15$ in clusters at the same radius.

\begin{figure}
\centering
\includegraphics[width=8.3cm]{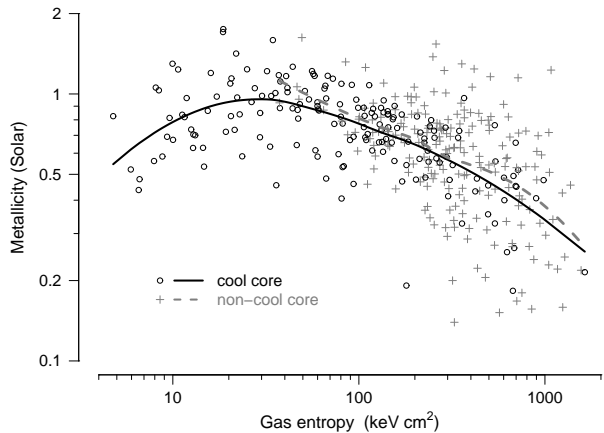}
\caption{ \label{fig:Z_vs_egas}
Projected gas metallicity vs. entropy, with locally weighted fits to the CC
and non-CC clusters plotted as solid and dashed lines, respectively.  }
\end{figure}

The similar abundance patterns between CC and non-CC clusters can be seen
by examining the gas metallicity as a function of entropy, plotted in
Fig.~\ref{fig:Z_vs_egas}. Also plotted are locally-weighted curves for the
combined points in each category, which demonstrate that there is very
little difference between them, aside from the greater radial coverage in
the centres of CC clusters. Since the gas entropy scales roughly linearly
with radius (Fig.~\ref{fig:S_vs_kpc}), the trend between metallicity and
entropy mimics that seen in Fig.~\ref{fig:Z(r)}. However, a striking aspect
of Fig.~\ref{fig:Z_vs_egas} is the almost complete absence of
low-metallicity gas (i.e. $<$ 0.4--0.5 Solar) with low entropy ($S<$
200\keV\cmsq). In general it appears that the most enriched gas has low
entropy, although the smoothed local regression suggests a turnover in this
trend towards lower metallicity below $\sim$30 \keV\cmsq. Interestingly,
this entropy level corresponds to the threshold below which star formation
appears to take place in the central galaxies of CC clusters
\citep{voit08,rafferty08}. Thus, a reversal of the inverse trend
between entropy and metallicity may reflect the loss of the most enriched
gas from the hot phase in fuelling such star formation.

\section{Cluster bimodality and thermal conduction}
\label{sec:bimodality}
Clusters can be divided according to the presence or absence of a cool
core, and it is clear that the properties of these two categories differ
substantially in terms of their temperature, density and entropy
profiles. However, an important question to ask is, are these two types
merely separate parts of the same continuous distribution, or do they
really constitute distinct populations? If the latter holds true, then this
would have interesting implications for models of self-similarity
breaking via feedback and/or other non-gravitational processes.

\begin{figure}
\centering
\includegraphics[width=8.4cm]{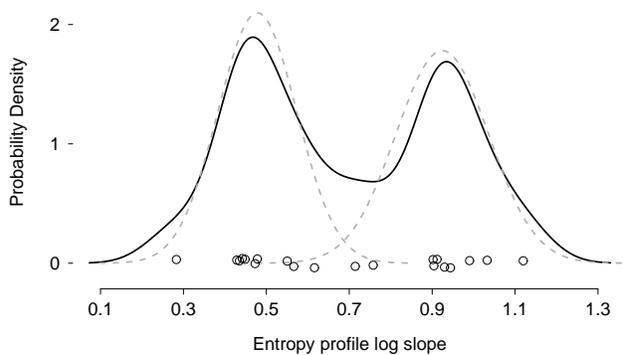}
\caption{ \label{fig:S(r)_index}
Kernel-smoothed (with a Gaussian of standard deviation 0.07) probability
density plot of the best-fit logarithmic slope of the entropy profile
(solid line), showing the raw values as randomly `jittered' points over the
X-axis. The dashed curves are the components of the best-fit bimodal
Gaussian distribution to the data (see text for details). With the
exception of Abell~2142, all the clusters above 0.7 are CC and all those
below are non-CC.}
\end{figure}

To address this issue, we return to the power-law fits to the entropy
profiles described in Section~\ref{ssec:S(r)_fit}. The probability
distribution of logarithmic slopes is plotted as a kernel density estimate
in Fig.~\ref{fig:S(r)_index}, with the positions of the raw values also
indicated. It can be seen that two peaks are visible in the distribution,
suggestive of bimodality in the cluster population. We performed a maximum
likelihood fit to the unbinned entropy slope values with both a single and
a pair of (equally-weighted) Gaussians, using the \textsc{fitdistr}
function in the \textsc{mass} package in \Rproject. The single Gaussian
best-fit mean value was $\mu=0.70\pm0.05$ with standard deviation of
$\sigma=0.25\pm0.04$, and the best fit double-Gaussian values were:
\begin{description}
\item $\mu1=0.48\pm0.04$~~~~$\sigma1=0.10\pm0.03$
\item $\mu2=0.92\pm0.04$~~~~$\sigma2=0.11\pm0.03$. 
\end{description}
The two separate Gaussians are plotted as dashed curves in
Fig.~\ref{fig:S(r)_index} and align closely with the peaks in the smoothed
density distribution; the cross-over point between these two Gaussians is
0.69 and is plotted as the dashed horizontal line in
Fig.~\ref{fig:S-index_vs_kT}.

In order to provide a quantitative assessment of the putative bimodality in
the population, we calculated both the Akaike information criterion
\citep[AIC,][]{akaike74} and the Bayesian information criterion
\citep[BIC,][]{schwarz78}. These are standard statistics used in model 
selection \citep[e.g. see][for a discussion of their relative
merits]{liddle07}, based on the negative log-likelihood penalised according
to the number of free parameters, with the model having the lowest value
always being preferred. The corresponding values obtained are AIC\,=\,4.65
(-0.49) and BIC\,=\,6.64 (3.49) for the single (double) Gaussian model,
implying a change in BIC of 4.0 in favour of the two component fit.
Differences in BIC of between 2 and 6 indicate positive evidence against
the model with the greater BIC value, with values above 6 indicating strong
evidence against the model with the greater BIC value
\citep[e.g.][]{mukherjee98}, demonstrating that a bimodal distribution is
clearly favoured over a unimodal distribution. Given that the sample was
statistically selected, this therefore implies that two distinct categories
of cluster exist.

The two separate distributions of entropy profile logarithmic slopes match
the cool-core classification (with the exception of Abell~2142; see
Section~\ref{ssec:S(r)_fit}), show similar dispersion ($\sigma\sim0.1$) and
are well-separated (the means differ by $\sim$4 standard deviations). The
means of the distributions are also in good agreement with the results
obtained by fitting Equation~\ref{eqn:S(r)_fit} to the CC and non-CC
clusters separately, as summarised in table~\ref{tab:S(r)_fit}.

While a power law fit describes the entropy profiles of CC clusters quite
well, it can be seen from Fig.~\ref{fig:sc065S_vs_r500} that the non-CC
profiles appear to flatten increasingly at small radii. This raises the
possibility that estimating the gradient from a power law fit (albeit
robustly) could bias the results. Therefore, to check our conclusions
regarding bimodality, we have also evaluated the entropy profile
logarithmic slope at 0.05\rfiveh\ as estimated from a locally-weighted fit
to the data (in log-log space), using the \textsc{loess} task in \Rproject\
(c.f. Fig.~\ref{fig:sc065S_vs_r500}), with a heavier smoothing (using a
value of the `span' parameter of 2), to increase the stability of the
gradient measurements. Repeating the above analysis leads to best fit
bimodal populations, given by
\begin{description}
\item $\mu1^{\prime}=0.26\pm0.09$~~~~$\sigma1^{\prime}=0.28\pm0.07$
\item $\mu2^{\prime}=0.97\pm0.03$~~~~$\sigma2^{\prime}=0.08\pm0.02$. 
\end{description}
The corresponding values obtained are AIC\,=\,25.4 (18.4) and BIC\,=\,27.4
(22.4) for the single (double) Gaussian model, implying a change in BIC of
5.0 in favour of the two component fit. It can be seen that the non-CC
slope is indeed flatter at this smaller radius (although with much greater
dispersion), but that the conclusion that the population is bimodal is
verified.

\subsection{Conduction and thermal balance}
One of the most intriguing aspects of the ICM is its resistance to runaway
radiative cooling. Although galaxy feedback is a plausible mechanism for
maintaining (or nearly maintaining) thermal balance in the centres of
cool-core clusters, it is less clear if it can similarly affect non-CC
clusters, which none the less also have short cooling times in their inner
regions \citepalias{san06}. This is because non-CC clusters show no sign of
mass drop-out necessary to fuel AGN outbursts or supernova winds. However,
heat transport by thermal conduction is likely to play an important role in
such cases and is also very effective at stabilizing cooling within CC
regions, provided the gas density is not too high
\citep[e.g.][]{conroy08,guo08}.

Circumstantial evidence for the effectiveness of thermal conduction in
counteracting cooling can be seen in the temperature profiles of CC
clusters. As pointed out by \citet{voigt02}, cool cores at the limit of
stablization by conduction would have temperature gradients of the form
$T\propto r^{0.4}$ in the case of bremsstrahlung emission, flattening to
$T\propto r^{0.3}$ for line-dominated emission, which concurs well with
observations of CC clusters \citepalias{san06} and groups (O'Sullivan et
al. in prep.), respectively. Furthermore, recent work by \citet{voit08} and
\citet{rafferty08} indicates that star formation in the cores of clusters,
resulting from unchecked cooling, only occurs when the gas entropy drops
below a certain level ($\la$\,30\keV\cmsq) which matches the threshold
below which gas becomes thermally unstable against conduction.

\begin{figure}
\centering
\includegraphics[width=8.4cm]{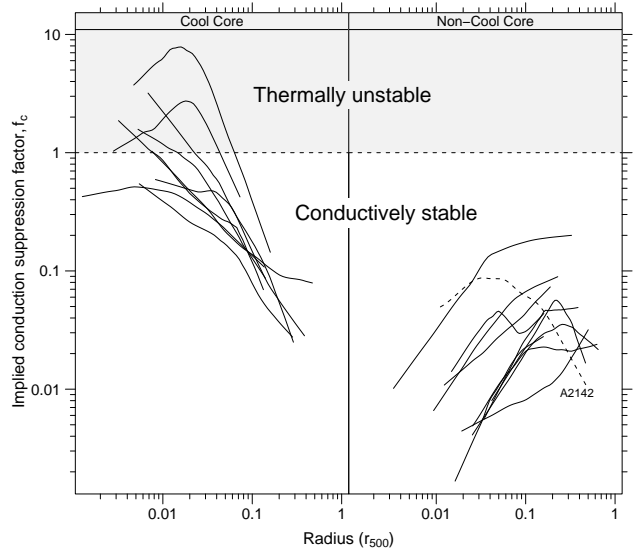}
\caption{ \label{fig:f.c(r)}
Implied Spitzer conductivity suppression factor as a function of scaled
radius, grouped by cool-core type and plotted with identical axes. Each
curve represents a locally weighted fit to the data points, to suppress
small-scale fluctuations (see the text for details).}
\end{figure}

The critical scale for thermal conduction is given by the Field length,
which must exceed the characteristic size of the system in order to allow
thermal balance to be maintained by smoothing out temperature fluctuations
\citep{field65}. Following \citet{donahue05} and \citet{voit08}, the Field
length, $\lambda_{\mathrm{F}}$, can be approximated as
\begin{equation}
\label{eqn:f.c}
\lambda_{\mathrm{F}} = \left( \frac{\kappa T}{\rho_{\mathrm{gas}}^2 \Lambda} \right)^{1/2}
\approx 4\,\kpc \left( \frac{S}{10\keV\cmsq} \right)^{3/2} f_c^{1/2}~,
\end{equation}
where $\kappa$ is the Spitzer conduction coefficient with suppression
factor $f_c$. This relation applies for the case of bremsstrahlung
emission, where the cooling function varies with temperature, $T$, such
that $\Lambda\propto T^{1/2}$, which renders $\lambda_{\mathrm{F}}$ a
function of entropy only \citep{donahue05}.

In the limit of conductive thermal balance, where $r=\lambda_{\mathrm{F}}$,
Equation~\ref{eqn:f.c} can be rearranged to yield an expression for the
corresponding implied suppression factor:
\begin{equation}
f_c \approx 62.5\, \frac{r^2}{S^3}~,
\end{equation}
in terms of the radius, $r$ (in kpc) and the entropy, $S$ (in
keV\cmsq). The variation of this quantity with scaled radius is plotted for
each cluster in Fig.~\ref{fig:f.c(r)}, smoothed using the \textsc{loess}
function in \Rproject\ and split by cool core status. Also shown is the
threshold of thermal stability, corresponding to conduction at the Spitzer
rate, when $f_c=1$. The striking aspect of Fig.~\ref{fig:f.c(r)} is the
clear difference between the profiles of CC and non-CC clusters, in terms
of both their shapes and locations on the plot: CC clusters having higher
normalization and largely negative gradients compared to non-CC profiles,
which all lie well within the conductively stable region with mostly
positive gradients. Once again, the outlier cluster Abell~2142 (as seen in
Fig.~\ref{fig:S-index_vs_kT}) stands out as an anomaly (the only non-CC
curve aligned from bottom right to top left), which nevertheless retains
similarities to both categories in that it lies within the range of the
\emph{non}-CC profiles while bearing closer resemblance to the CC profiles in
shape.

The implication of Fig.~\ref{fig:f.c(r)} is that non-CC clusters are
certainly capable of being stablilized by conduction alone, operating at no
more than $\sim$10 per cent of the Spitzer rate and even much less
effectively than that within their inner regions. By contrast, the CC
clusters require values of $f_c\ga 0.1$ within the peak temperature radius
\citepalias[$\sim0.15$\rfiveh;][]{san06}, rising sharply towards the 
centre and in some cases crossing into the region of thermal instability,
corresponding to $f_c>1$ (for NGC~5044, Abell~262, Abell~478, Abell~496 and
2A~0335+096). For these five clusters, an additional heat source would be
required in order to maintain thermal balance. With the exception of A478,
these are 4 of the 5 coolest CC clusters (see Table~\ref{tab:main}), which
is unsurprising, since thermal conduction operates much less effectively at
lower temperatures as the Spitzer conductivity, $\kappa\propto T^{5/2}$.
Furthermore, all except Abell~496 show evidence of AGN-related disturbance
in the form of X-ray cavities \citep{birzan04} and both X-ray and radio
disturbance in the case of NGC~5044 (David et al, in prep.). These results
are in agreement with the findings of \citet{voigt04}, who studied four of
the same clusters: for Abell~2029 \& Abell~1795 they also concluded that
conductivity at or below the Spitzer level was able to balance cooling
everywhere inside the cooling radius, whereas this was not the case for
Abell~478 and 2A~0335+096.

Theoretical considerations indicate that magnetohydrodynamic turbulence in
the ICM could give rise to conduction suppression factors of $\sim$0.2
\citep{narayan01}, which is also consistent with results from the direct
numerical simulations of \citet{maron04}, who favour $f_c\sim$
0.1--0.2. This would imply conductive stability for the non-CC clusters at
all radii and for all CC clusters outside $\sim$0.05--0.1\rfiveh\
(Fig.~\ref{fig:f.c(r)}). Taking all the $f_c$ values for the non-CC
clusters, we find the spread of values to be well fitted by lognormal
distribution with a mean log (base 10) of $-1.50\pm0.03$ and log standard
deviation of $0.39\pm0.02$ corresponding to 0.032 with a 1$\sigma$ range of
0.013--0.077.

Why, though, should a threshold for conduction stability give rise to a
well-separated bimodal distribution of clusters, with a `zone of avoidance'
in between? \citet{donahue05} point out that clusters exceeding the
threshold for conductive stability will continue to cool until some other
form of heating intervenes, which would therefore lead to diverging
populations. In which case, the fact that the non-CC $f_c$ profiles are
themselves not flat suggests that this threshold is spatially varying.
Since the likely dominant cause of conduction suppression is magnetic
fields, it follows that the configuration of magnetic field lines varies
significatly with radius. Consequently, it is possible that the variation
of $f_c$ profiles amongst the non-CC clusters reflects the intrinsic
variation in intracluster magnetic field configurations in the cluster
population. In any case, the non-CC $f_c$ profiles (except for A2142) all
drop sharply inside $\sim$0.1\rfiveh, reaching values within the typical
scale of the central cluster galaxy ($\sim$0.02\rfiveh) consistent with the
higher suppression factors of $\ga$\,0.01 present in the interstellar
medium of cluster galaxy coronae \citep{sun07}. This indicates that the
transition within the ICM to higher suppression factors on galaxy scales
may occur smoothly, suggesting a gradual variation in magnetic fields.

By contrast, the presence of cold fronts in the ICM implies sharply
discontinuous magnetic fields, especially considering that such features
occur at larger cluster radii, where $f_c$ is greater. For example, in the
case of Abell~2142, detailed analysis of the temperature gradient across
the cold front indicates that conduction must be suppressed by at a factor
of 250--2500 \citep{ettori00}, which is certainly far below the levels
implied in Fig.~\ref{fig:f.c(r)}, although these amount to
globally-averaged estimates.  Such discontinuous configurations are
possible in this case, since the magnetic field is likely to be stretched
by tangental gas motions near the cold front, making it stronger and
changing its structure compared to the rest of the ICM
\citep{vikhlinin01b}.

A notable feature of Fig.~\ref{fig:f.c(r)} is the absence of any
substantially flat profiles, in between the strongly negative gradients of
the CC clusters and the mostly strongly positive (except for A2142)
gradients of the non-CC clusters. A flat profile, lying along the
conduction-stabilized threshold with a constant value of $f_c\le 1$, would
imply an entropy profile of the form $S\propto r^{2/3}$ \citep{donahue05},
which is close to the cross-over point of 0.69 between the two
distributions of entropy profile logartithic slopes found in
Section~\ref{sec:bimodality} (also plotted as the dashed horizontal line in
Fig.~\ref{fig:S-index_vs_kT}). Thus the divergence between CC and non-CC
profiles seen in Fig.~\ref{fig:f.c(r)} mirrors the bimodality seen in the
slopes of the entropy profiles described above.

\section{Discussion}
\label{sec:discuss}
The above results clearly indicate that bimodality is present in the
cluster population, structured along the dichotomy between cool core and
non-cool core clusters. Furthermore, the implication is that thermal
conduction alone is a plausible mechanism for stabilizing the intracluster
medium in non-CC clusters, whereas additional heating from galaxy feedback
is necessary to achieve the same in CC clusters-- a conclusion also reached
by \citet{guo08}. Conductive heat transfer could certainly explain why no 
significant temperature decline is observed in non-CC clusters, despite the
short cooling times of gas in their cores \citepalias{san06}. However, by
contributing to the heating of the cluster core, conduction also acts to
reduce the amount of energy input required from additional sources such as
AGN, in order to maintain thermal stability. This may therefore account for
the observed lack of evidence for \emph{strong} AGN heating in clusters
\citep{mcnamara07}.

Nevertheless, while feedback heating may not be necessary to stabilize
non-CC clusters, it is clear from the results of the entropy-temperature
scaling analysis in Section~\ref{ssec:S_scaling} that both CC and non-CC
clusters show significant departures from self-similarity. Such similarity
breaking is the unmistakable signature of non-gravitational physics, which
suggests that the non-CC clusters must \emph{also} have been impacted
significantly by feedback (if not also radiative cooling) in their
lifetimes. This possibility is consistent with the theoretical model of
\citet{mccarthy08}, where non-CC clusters are formed from material that
has experienced higher levels of preheating. In this picture, the influence
of conduction could help to segregate the cluster population and stave off
the formation of cool cores in the most strongly preheated systems.

A related aspect is the role of mergers in cluster evolution, particularly
given the close association between signatures of recent disruption, such
as radio halos, and the absence of a cool core
\citep[e.g.][]{million09}. Although suggesting a key role for 
thermal conduction in sustaining non-CC clusters, our results nevertheless
certainly do not rule out merging as an alternative explanation for their
existence. While recent simulations have concluded that cluster mergers
cannot permanently erase cool cores \citep{poole06}, it is possible that
the additional influence of conduction could achieve this outcome and
thereby provide an alternative path to a non-CC state.  In this situation,
the temporary fragmentation of a cool core that can take place in a merger
\citep{poole08} could plausibly lead to conductive dissipation of the
resulting blobs of cool gas, whose size can easily fall below the Field
length, provided the magnetic field configuration is favourable. Since
conduction lowers the threshold necessary to transition from a state where
additional feedback is required (i.e. in a cool core), to one where
conduction alone can maintain stability (i.e. a non-cool core), the
transformation from CC to non-CC status is correspondingly more achievable.

However, notwithstanding this appealing explanation of bimodal populations,
it is also clear that the key role that conduction can play in stabilizing
\emph{clusters} against cooling instabilities cannot easily extend to galaxy
groups, where the lower gas temperatures render it much less effective (the
conduction coefficient, $\kappa \propto T^{5/2}$). Therefore, the impact of
merging activity may be more short lived in galaxy groups, without the
contribution of significant conductive heat transfer to impede re-formation
of a cool core. This suggests that non-cool core groups are likely to have
been \emph{recently} disrupted; by contrast, conduction in clusters could
sustain a non-CC state in post mergers long after disruption. Nevertheless,
it remains to be seen whether a consistent thermodynamic picture of ICM
evolution in both groups \emph{and} clusters can be developed.

The similarity of the ICM metallicity in CC and non-CC clusters is
noteworthy as the only property of the gas that does not differ
significantly between the two types, at least within the region of overlap
seen in this sample. This fact alone provides evidence that strong mixing
of gas cannot preferentially have affected non-CC clusters-- as a result of
merging activity, for example. However, the width of the metallicity peak
in the cluster cores appears significantly ($\ga2\times$) broader than the
stellar distribution of the brightest cluster galaxy (BCG), as previously
noted by \citet{rebusco05}, which implies some mechanism to transport
metals into the ICM. \citeauthor{rebusco05} speculate that this might be
caused by AGN-driven gas motion, as also favoured by the recent theoretical
modelling of \citet{rasera08}. However, if AGN are responsible for
diffusing the enriched gas then the similarity of cluster metallicity
profiles implies an equally prominent role for AGN outflows in non-CC
clusters, despite the lack of evidence for significant AGN disruption in 
such cases.

\section{Conclusions}
Using the statistically-selected sample of 20 galaxy clusters presented in
\citet{san06}, we have studied the density, entropy and metallicity of the
intracluster medium as a function of radius, focussing on the comparison
between clusters with cool-cores (CC) and those without. We describe an
improved method of estimating the cluster mean temperature and fiducial
scaling radius \rfiveh, which we use to explore systematic trends in
cluster gas properties across the sample.

We find that the gas density is systematically higher in the cores of CC
clusters, and that the ICM is progressively depleted in less massive
systems. We also find a clear departure from self-similar scaling in the
gas entropy which is consistent with the modified scaling of $S\propto
\Tbar^{b}$ with $b=0.65$ from \citet{pon03}: at 0.1\rfiveh\ the best-fit scaling is $b = 0.66\pm0.10$ and
$0.71\pm0.21$ for CC and non-CC clusters, respectively. However, the
dependency on temperature strenghens when all the clusters are combined, to
a nearly self similar value of $b = 0.92\pm0.12$, and similar results are
obtained in all three cases for entropy measured at 0.15\rfiveh\ versus
mean temperature. This demonstrates that similarity breaking
(i.e. $S\propto T^{\sim2/3}$) can exist in the separate populations of CC
and non-CC clusters, even while the combined population shows consistency
with self-similarity.

The metallicity of the gas shows no evidence of a systematic variation with
\Tbar, but declines with radius such that $Z\propto r^{-0.31\pm0.04}$ 
outside 0.014\rfiveh\ (comparable to the size of any central dominant
galaxy), for both CC and non-CC clusters alike. Inside this point there is
substantial divergence in the metallicity, with a few CC clusters showing
sharply decreasing $Z$ towards the centre while others possess continually
rising profiles. At large radii, there is no indication of any flattening
in the metallicity profile to at least $\sim$0.5\rfiveh, where the
\Chandra\ field-of-view limits the data. We study gas metallicity as a
function of entropy and find a striking lack of low-metallicity gas
(i.e. $<$ 0.4--0.5 Solar) with low entropy ($S<$ 200\keV\cmsq). Above
$\sim$100\keV\cmsq\ the metallicity declines with increasing entropy in an
identical fashion for both CC and non-CC clusters.

We address the issue of bimodality in cluster properties by studing the
distribution of logarithmic slopes obtained from power-law fits to
individual cluster entropy profiles (i.e. $S\propto r^a$). We find that a
double-Gaussian distribution is strongly preferred over a unimodal Gaussian
distribution, using maximum likelihood fits to the unbinned values,
employing both the Bayesian and Akaike Information Criteria model selection
tests. The best fitting means of the two distributions are $a=0.92\pm0.04$
and $0.48\pm0.04$, with a standard deviations of 0.1 in both cases. Given
the statistically selected nature of the sample, this demonstrates that two
distinct categories of cluster exist, which has important implications for
models of galaxy feedback and cluster similarity breaking.

We explore the impact of thermal conduction on the ICM by studying the
implied conduction suppression factor, $f_c$, as a function of radius. We
find that the profiles of $f_c$ differ sharply between CC and non-CC
clusters consistent with two distinct populations. We conclude that
conduction alone is capable of stabilizing non-CC clusters against
catastrophic cooling, while in CC clusters some feedback is required in
addition to conduction to maintain thermal balance, in agreement with the
findings of \citet{guo08}. Taking all the $f_c$ values for the non-CC
clusters, we find the spread of values to be well fitted by lognormal
distribution with a mean log (base 10) of $-1.50\pm0.03$ and log standard
deviation of $0.39\pm0.02$ corresponding to 0.032 with a 1$\sigma$ range of
0.013--0.077.

\section*{Acknowledgments}
AJRS thanks Ria Johnson for useful discussions and for porting the BCES
regression code to \Rproject. We thank the referee, Megan Donahue, for
useful comments which have improved the clarity of the paper. AJRS
acknowledges support from STFC, and EOS acknowledges support from NASA
awards AR4-5012X and NNX07AQ24G. This work made use of the NASA/IPAC
Extragalactic Database (NED) and the \Rproject\ tutorials at the Penn State
Center for Astrostatistics.

\bibliography{/data/ajrs/stuff/latex/ajrs_bibtex}
\label{lastpage}

\end{document}